%% file: mentoring.tex
\documentclass[acmsmall]{acmart}
\usepackage{graphicx}
\usepackage{pgfplots}

\AtBeginDocument{%
  }

\setcopyright{acmcopyright}
\copyrightyear{2018}
\acmYear{2018}
\acmDOI{XXXXXXX.XXXXXXX}

\acmConference[Conference acronym 'XX]{Make sure to enter the correct
  conference title from your rights confirmation emai}{June 03--05,
  2018}{Woodstock, NY}
\acmPrice{15.00}
\acmISBN{978-1-4503-XXXX-X/18/06}

\begin{document}

\newcommand{\new}[1]{#1}

\title{Long-Term Mentoring for Computer Science Researchers}

\author{Emily Ruppel\**}
\affiliation{%
  \institution{Carnegie Mellon University}
  \country{USA}}

\author{Sihang Liu\**}
\affiliation{%
  \institution{University of Virginia}
  \country{USA}}
\thanks{\** Co-first authors \& CALM co-chairs}

\author{Elba Garza}
\affiliation{%
  \institution{Texas A\&M University}
  \country{USA}}

\author{Sukyoung Ryu}
\affiliation{%
 \institution{KAIST}
 \country{Korea}}

\author{Alexandra Silva}
\affiliation{%
  \institution{Cornell University}
  \country{USA}}

\author{Talia Ringer\*$^{\diamond}$}
\affiliation{%
  \institution{University of Illinois Urbana-Champaign}
  \country{USA}}
\thanks{\*$^{\diamond}$ Last author \& SIGPLAN-M chair}





\maketitle


\input{background}
\input{design}

\input{impact}
\input{challenges}
\input{forward}

\begin{acks}
\new{
We would like to thank the current and past committees of SIGPLAN-M and CALM, along with
all of the mentors and mentees in both programs, without whom none of this would be possible.
We are grateful to everyone who has given us feedback on our programs over the last two years,
and to everyone who gave us feedback on this article.
And we will forever remember the unwavering support from the many senior leaders
in our research communities who empowered us to lead ourselves.
Thank you.}
\end{acks}

\bibliographystyle{ACM-Reference-Format}
\bibliography{mentoring}

\input{appendix}

\end{document}

%% file: background.tex
Early in the pandemic, we---leaders in the research areas 
of programming languages (PL) and computer architecture (CA)---realized that we had a problem:
the only way to form \textit{new} lasting connections in the community was to 
\textit{already have} lasting connections in the community.
Both of our academic communities had wonderful short-term mentoring programs 
to address this problem, but it was clear that we needed
long-term mentoring programs.

Those of us in CA approached this scientifically,
making an evidence-backed case for community-wide long-term mentoring~\cite{Garza2021WCAE}.
In the meantime, one of us in PL had impulsively
launched an unofficial long-term mentoring program,
founded on chaos and spreadsheets.
In January 2021, the latter grew to an official cross-institutional
long-term mentoring program called SIGPLAN-M;
in January 2022, the former grew to 
\underline{C}omputer \underline{A}rchitecture \underline{L}ong-term \underline{M}entoring (CALM).

The impacts have been strong: SIGPLAN-M reaches 328 mentees and 
234 mentors across 41 countries, and mentees have described
it as ``life changing'' and ``a career saver.''
And while CALM is in its pilot phase---with 13 mentors and 21 mentees across 7 countries---it
has received very positive feedback. 
The leaders of SIGPLAN-M and CALM
shared our designs, impacts, and challenges along the way.
Now, we wish to share those with you.
We hope this will kick-start
a larger long-term mentoring effort across all of computer science.

%% file: design.tex
\section{Designing a Long-term Mentoring Program}

We designed SIGPLAN-M and CALM to address two gaps in our communities:

\begin{enumerate}
    \item helping junior and aspiring researchers form long-term connections in our communities, and access the perspectives of researchers from other institutions (CALM and SIGPLAN-M), and
    \item helping senior researchers access mentorship of any kind (SIGPLAN-M only).
\end{enumerate}

\paragraph{Organization} Both SIGPLAN-M and CALM are run by volunteers. Both programs have an
\textbf{operations committee} of junior researchers who handle
matching, recruitment, and other operational tasks.
SIGPLAN-M also has an \textbf{advisory committee}
of senior researchers who help communicate with leadership in SIGPLAN and the ACM.
CALM is developing a similar advisory committee, which is especially useful given that CA spans both the ACM and IEEE.

\begin{figure}
\begin{minipage}{.45\textwidth}
\centering
\footnotesize
\begin{tabular}{r|c|c}
  & Mentees & Mentors \\
  \hline
  High School Students & 3 & 0 \\
  \hline
  Undergraduate Students & 53 & 0 \\
  \hline
  Masters Students & 35 & 1 \\
  \hline
  Ph.D. Students & 135 & 34 \\
  \hline
  Software Engineers & 43 & 11 \\
  \hline  
  Post-Doctoral Researchers & 8 & 24 \\
  \hline
  Government Researchers & 0 & 6 \\
  \hline
  Industrial Researchers & 9 & 36 \\
  \hline
  Industrial Executives & 3 & 5 \\
  \hline
  Teaching Faculty & 2 & 4 \\
  \hline
  Research Faculty & 2 & 5 \\
  \hline
  Pre-Tenure Faculty & 12 & 40 \\
  \hline
  Tenured Faculty & 2 & 59 \\
  \hline
  Other or Unknown & 21 & 9 \\
  \hline
  Total & 328 & 234 \\
\end{tabular}
\end{minipage}
\hfill
\begin{minipage}{.45\textwidth}
    \centering
    \includegraphics[scale=0.27]{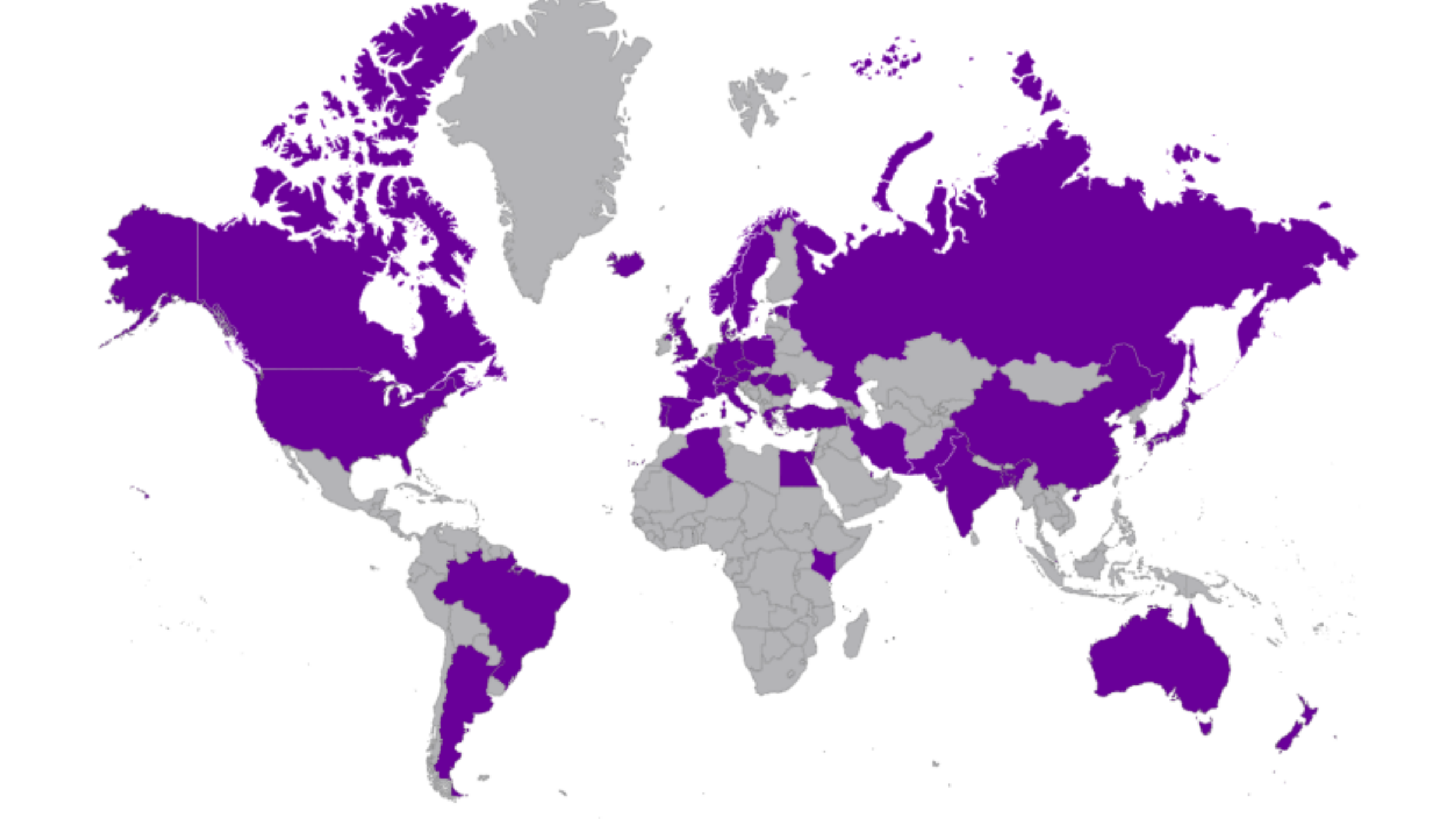}
\end{minipage}
\caption{Mentors and mentees in SIGPLAN-M as of July 2022 (left), and map of where they live (right).}
\label{fig:summary}
\end{figure}

\paragraph{Scope} SIGPLAN-M is open to \textbf{any seniority level}
in \textbf{any country}---see
Figure~\ref{fig:summary} for a current breakdown.
It is possible to serve as both a mentee and a mentor at the same time
\new{(this is common, and helps with mentor recruitment)}.
CALM is similarly global, but is piloting for \textbf{students}.
The scope of mentoring can be any mix of \textbf{technical} and
\textbf{non-technical (career)} topics,
including the experiences 
of \textbf{historically marginalized groups} in computing.

\paragraph{Recruitment} Both CALM and SIGPLAN-M recruit mentors and mentees \textbf{in batches before conferences}.
\new{This makes it possible for us to reuse conference registration infrastructure, and to piggyback
off of existing outreach and recruitment efforts for colocated short-term mentoring workshops.}
SIGPLAN-M also recruits \textbf{off-cycle on a rolling basis} via registration forms on our website,
and using social media, flyers, stickers,
and presentations at major conference business meetings.
We occasionally target mentor recruitment toward particular needs.

\paragraph{Registration} The registration forms ask participants their motivations, topics of interest,
and topic priorities.
Their \textbf{open-ended} questions allow for both flexibility in answers and vetting of participants.
The SIGPLAN-M forms also provide \textbf{example topics},
including some that may be taboo (like mental health).
They also include fields for preferred matches and matches to avoid.

\paragraph{Matching} We form matches based on registration data\new{, using guidelines discussed and revised in committee meetings}.
We deliberately form \textbf{cross-institutional} matches.
If no match is available, we \textbf{waitlist} mentees and revisit.
After matching, we email the mentor and mentee to
\textbf{initiate the relationship}\new{, using a set of shared email templates}.

\paragraph{Mentorship} SIGPLAN-M provides a mentoring guide~\cite{guidelines},
which advises mentors and mentees to focus the first conversation on \textbf{defining the relationship},
and on norms of \textbf{communication} and \textbf{confidentiality}.
\new{Communication frequency and medium are among the norms negotiated:
a typical commitment is one 30 minute video chat per month,
but details vary by match, and some matches communicate only as needed.}
Both CALM and SIGPLAN-M send \textbf{check-in emails} every two months to help participants navigate mentoring relationships and address any issues.

\paragraph{Renewal}
For both SIGPLAN-M and CALM, the default relationship is \textbf{one year} long\new{---long enough
to establish common ground, but short enough to provide an easy out}.
After a year, SIGPLAN-M asks participants
if they would like to \textbf{renew} the match,
\textbf{rematch} with someone else, or \textbf{withdraw}.
It also allows early withdrawal and rematching, if requested.

%% file: impact.tex
\section{Impacts on our Communities}
\label{sec:impacts}

\paragraph{SIGPLAN-M}
Currently, we have 328 mentees and 234 mentors
spanning 41 countries (Figure~\ref{fig:summary}).
After one year, we ran a survey to gather feedback; a full summary is in Appendix~\ref{sec:sigm-survey}.
Among respondents (67 mentees and 51 mentors),
\textbf{satisfaction} (1-5) was very high for mentees
(median 5, mean 4.43, standard deviation 0.94),
and slightly lower for mentors
(median 4, mean 4.12, standard deviation 0.89).
We observed a gap between how much \textbf{benefit} (1-5) mentees reported (median 5, mean 4.12, standard deviation 1.21), and how much mentors perceived their mentees as
benefiting (median 4, mean 3.51, standard deviation 0.85).
We responded to this by better communicating mentor impacts.
Highly satisfied participants cited common backgrounds or interests,
good communication, kindness, and helpful advice;
unsatisfied participants cited poor communication.

According to the feedback, SIGPLAN-M has helped mentees without access to local expertise
build bridges in the community, and has also had a strong diversity impact.
SIGPLAN-M has been particularly successful at pairing transgender mentees with transgender mentors
in PL---a need we had not anticipated, but that we are happy to meet.
Other outcomes have included help securing PhD positions or jobs,
recognizing and leaving unhealthy environments,
and forming international connections.
\new{While it is still too early to monitor long-term outcomes,}
we are thankful to our mentors for making such a big difference
in mentees' lives \new{already}.

\paragraph{CALM}
In our pilot, we have paired 21 mentees with 13 mentors
across 7 countries.
Mentees choose from two mentoring ``tracks'': 1) research and 2) personal development.
Most mentee applicants (64.5\%) preferred the research track.
While the original vision was for these tracks to remain separate, in practice the track selection was used loosely for forming better matches.

Initial qualitative feedback was largely positive,
and like the SIGPLAN-M feedback, demonstrated increased
opportunities for students without access to local experts.
CALM participants sometimes had difficulty establishing contact with their matches, and expressed concern that more communication is required to educate mentees on how to make the most of a mentorship.
CALM is exploring optional communication channels beyond email
to address communication gaps, and is reviewing its onboarding
process to better educate mentors and mentees.

%% file: challenges.tex
\section{Challenges we wish we'd known}



\paragraph{Workload} Running a long-term mentoring program is a \textit{lot} of work.
The SIGPLAN-M chair spends about five hours per week on this;
the CALM co-chairs spend about two hours per week each.
Community work of this kind---while massively impactful---remains
systemically unrewarded at hiring, promotion, and tenure time. 
This must change.

\textbf{Challenges:} Matching is hard work,
and there is a lot of maintenance that follows:
check-ins, rematching, renewals, and a never-ending waitlist.

\textbf{So far:} To better motivate volunteers, we have set
concrete days for tasks, ensured clear ownership over tasks,
and paired committee members to help newcomers learn the ropes.

\textbf{Next year:} We plan to set more concrete
roles for volunteers. 
In lieu of concrete roles for tasks like check-ins and renewals,
these tasks often fall on the chairs.

\textbf{Wishes:} Work serving our professional communities
needs to be systemically rewarded.

\paragraph{Infrastructure} Both programs are still run very manually, which is time-consuming, and has led to unexpected challenges like
being marked as spammers by some email clients.

\textbf{Challenges:} We do not want a fully automated matching process,
as human attention to matches is important.
The infrastructure that we want is nontrivial,
and potentially expensive.

\textbf{So far:} We have created email templates,
and we have documented manual processes.
SIGPLAN-M has one programmer working on infrastructure, but progress has been slow.

\textbf{Next year:} We hope to set up simple email automation,
and appoint someone familiar with our needs to a dedicated role
managing the development of our infrastructure.

\textbf{Wishes:}
We need infrastructure for searching and filtering potential matches,
managing mentor and mentee profiles, keeping track of matches and capacity, and automating email.
The best path could be to pool resources and build common
infrastructure.
More support from professional societies and buy-in from
other research areas would help.

\paragraph{Support} Each of our committees is a handful of volunteers managing tens to hundreds of mentors and mentees.
The mentors and mentees could use a lot more support.

\textbf{Challenges:} Mentors tend to overcommit
or lose track of communication.
Recruiting mentors for specific needs can be hard.
Mentors and mentees often need coaching around skills like communication.
Mentees are sometimes poor fits for the program,
or change interests after starting.

\textbf{So far:} We provide coaching through check-ins. To motivate mentors, SIGPLAN-M acknowledges mentors on our website~\cite{website} on an opt-in basis, and highlights exceptional mentors on the SIGPLAN blog~\cite{blog1, blog2}. When SIGPLAN-M is a bad fit for a mentee,
we try to help them find a mentor elsewhere.

\textbf{Next year:} We hope to better clarify what makes a mentee a good fit. We also hope to appoint dedicated roles to help
with coaching mentors,
managing mentor-mentee relationships, motivating and rewarding mentors,
and recruiting mentors for specific needs.

\textbf{Wishes:} We want long-term mentoring programs in other communities, so that we can direct mentees toward other programs when appropriate.
This will also help with shared dedicated roles across programs, for things that rely little
on the details of particular research areas.

%% file: forward.tex
\section{Going Forward}


We are extremely grateful for such active engagement from
our communities. We hope to spread our models of long-term mentoring
beyond PL and CA,
to reach research communities all across computer science.
All it takes to get started is a handful of volunteers in your research community willing to put in the work.
If this is you, please contact us, and we will joyfully help you get started.

%% file: appendix.tex
\newpage

\appendix
\section*{Appendix A: SIGPLAN-M Survey Summary}
\label{sec:sigm-survey}

\begin{figure}
    \centering

\begin{minipage}{.48\textwidth}
\centering
    \begin{tikzpicture}
    
    \pgfplotsset{width=7 cm, height=4cm}
 
\begin{axis}[ybar, bar width=10pt, ymin=0, xtick=data,
    ylabel={\# Mentees},
    y label style={at={(axis description cs:0.1,.5)},anchor=south},
legend style={at={(0.25,0.95)},
 anchor=north}]
\addplot [fill=black] coordinates {
    (1,2) 
    (2,1) 
    (3,6) 
    (4,15)
    (5,43)
};
\addplot [fill=white] coordinates {
    (1,5) 
    (2,2) 
    (3,8) 
    (4,13)
    (5,39)
};
\legend{Satisfaction, Benefit}
\end{axis}

\end{tikzpicture}
\end{minipage}
\hfill
\begin{minipage}{.45\textwidth}
\centering
    \begin{tikzpicture}
    
    \pgfplotsset{width=7 cm, height=4cm}
 
\begin{axis}[ybar, bar width=10pt, ymin=0, xtick=data,
    ylabel={\# Mentors}, 
    y label style={at={(axis description cs:0.1,.5)},anchor=south},
legend style={at={(0.25,0.95)},
 anchor=north}]
\addplot [fill=black] coordinates {
    (1,1) 
    (2,0) 
    (3,11) 
    (4,18)
    (5,21)
};
\addplot [fill=white] coordinates {
    (1,1) 
    (2,5) 
    (3,16) 
    (4,25)
    (5,4)
};
\legend{Satisfaction, Benefit}
\end{axis}
 
\end{tikzpicture}
 \end{minipage}
 
    \caption{Quantitative survey results at the end of the first year of SIGPLAN-M, for the 67 mentees (left) and 51 mentors (right) who responded. Both mentors and mentees were asked how satisfied they were with their matches (black bars, scale of 1-5). Mentees were asked how much they benefited, while mentors were asked how much they thought their mentees benefited (white bars, scale of 1-5).}
    \label{fig:match}
\end{figure}
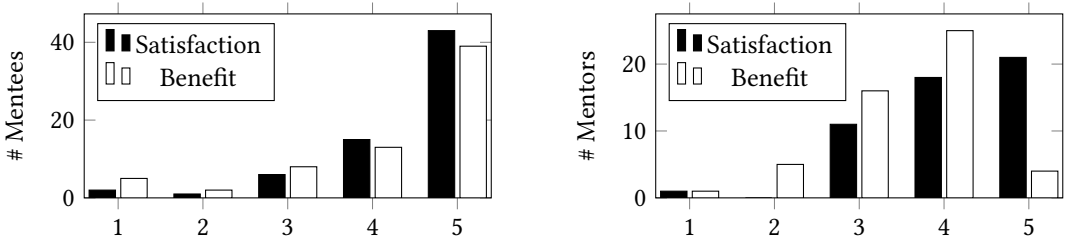

Figure~\ref{fig:match} summarizes the quantitative results for the SIGPLAN-M survey described in
Section~\ref{sec:impacts}. In our survey, we also asked mentors and mentees to share anecdotes, which we elaborate on
here, with informed consent.

\paragraph{What makes a good match?}
Highly satisfied mentees often noted that
their mentors gave helpful advice, shared common backgrounds or interests,
or were kind or good communicators:

\begin{quote}
[My] mentor understands me \ldots as he comes from the \textbf{same country} \ldots [he]
has picked me from zero knowledge \ldots to a point where I know the track ahead.
\end{quote}

\begin{quote}
She was \textbf{very approachable and sweet} and gave me some really \textbf{good advice}.
\end{quote}

\begin{quote}
He \textbf{always made time to meet me} when I needed it.
\end{quote}

\begin{quote}
My mentor is \textbf{an expert in the areas I wanted to learn more about} \ldots I've been able to [ask] questions about [career and personal topics] along with technical topics.
\end{quote}

Highly satisfied mentors noted similar themes:

\begin{quote}
My mentee has a \textbf{very similar background} \ldots which helps immensely.
\end{quote}

\begin{quote}
One of my mentees is \textbf{always in contact with me}, and we are discussing different aspects of PhD based on the mentee's experience, which is so cool.
\end{quote}

\begin{quote}
We are a good match \ldots and have \textbf{compatible research interests}.
\end{quote}

\begin{quote}
My mentee is \textbf{very responsive} and \textbf{not afraid to ask things from me}. 
\end{quote}

Consistent with this, both mentors and mentees who were unsatisfied with their matches
most commonly cited communication lapses:

\begin{quote}
My  mentor \textbf{never contacted me}.
\end{quote}

\begin{quote}
We did a few mail exchanges. But \textbf{the mentor seemed to drop the ball}. I don't think he was interested in the connection \ldots I feel let down as a mentee.
\end{quote}

\begin{quote}
[My mentor] must be busy and \textbf{has not contacted me}, and \ldots \textbf{I have not been able to gather the courage to contact her myself}.
\end{quote}

\begin{quote}
[My mentee] \textbf{stopped answering emails}.
\end{quote}

Of course, as one mentor hinted, it is sometimes hard to know whether poor communication
is the \textit{cause} of an unsatisfying match---and it may well be an \textit{effect}:

\begin{quote}
I think I was just \textbf{unable to connect to my mentees}, so \ldots they had no need to follow up (or didn't really like me as a mentor). 
\end{quote}
\new{It is also possible that mentor and mentee personalities may factor into communication lapses,
and may be worth considering both in matching and in our guidelines for establishing communication norms during the first meeting.}
More investigation into communication lapses would certainly help us improve our program!

\paragraph{What outcomes have we observed?}
SIGPLAN-M has helped mentees without access to local domain experts
build bridges with domain experts in the community:

\begin{quote}
[My mentor] also pointed me the directions which I could try, when \textbf{my university does not offer related opportunities} to the field which I am interested in.
\end{quote}

\begin{quote}
I wouldn't be able to \textbf{obtain technical feedback} \ldots if SIGPLAN-M did not exist.
\end{quote}

While we did not ask for consent to share quantitative diversity data,
we have anecdotal evidence of strong diversity impacts.
For example, transgender mentees have, in many cases, requested transgender mentors
to help navigate the challenges of being transgender in our research community:

\begin{quote}
I joined the program mostly for help \textbf{navigating academia and the job market while trans}, and I feel like I definitely have been getting help in this area!    
\end{quote}
We had not anticipated this before starting SIGPLAN-M, but it makes sense now---for transgender
mentees, it can be outright dangerous to put oneself out there searching for a transgender mentor.
SIGPLAN-M allows mentees to request a transgender mentor discreetly.

\new{We do not yet have an easy way of tracking long-term outcomes like retention for mentors and mentees.
Furthermore, it is sometimes unclear what outcomes it even makes sense to aim for---sometimes,
a good outcome for a mentee is the realization that they would rather do work in another field!
We believe strongly in the value of anecdotes for understanding successful outcomes may look like,
and what metrics we may wish to measure in aiming for those outcomes.}

Other outcomes are described in Section~\ref{sec:impacts},
along with a summary of our findings.

%% file: mentoring.bbl

\begin{thebibliography}{5}


\ifx \showCODEN    \undefined \def \showCODEN     #1{\unskip}     \fi
\ifx \showDOI      \undefined \def \showDOI       #1{#1}\fi
\ifx \showISBNx    \undefined \def \showISBNx     #1{\unskip}     \fi
\ifx \showISBNxiii \undefined \def \showISBNxiii  #1{\unskip}     \fi
\ifx \showISSN     \undefined \def \showISSN      #1{\unskip}     \fi
\ifx \showLCCN     \undefined \def \showLCCN      #1{\unskip}     \fi
\ifx \shownote     \undefined \def \shownote      #1{#1}          \fi
\ifx \showarticletitle \undefined \def \showarticletitle #1{#1}   \fi
\ifx \showURL      \undefined \def \showURL       {\relax}        \fi
\providecommand\bibfield[2]{#2}
\providecommand\bibinfo[2]{#2}
\providecommand\natexlab[1]{#1}
\providecommand\showeprint[2][]{arXiv:#2}

\bibitem[Garza et~al\mbox{.}(2021)]%
        {Garza2021WCAE}
\bibfield{author}{\bibinfo{person}{Elba Garza}, \bibinfo{person}{Gururaj
  Saileshwar}, \bibinfo{person}{Udit Gupta}, \bibinfo{person}{Tianyi Liu},
  \bibinfo{person}{Abdulrahman Mahmoud}, \bibinfo{person}{Saugata Ghose}, {and}
  \bibinfo{person}{Joel Emer}.} \bibinfo{year}{2021}\natexlab{}.
\newblock \showarticletitle{Mentoring Opportunities in Computer Architecture:
  Analyzing the Past to Develop the Future}. In \bibinfo{booktitle}{\emph{2021
  ACM/IEEE Workshop on Computer Architecture Education (WCAE)}}.
  \bibinfo{pages}{1--9}.
\newblock
\urldef\tempurl%
\url{https://doi.org/10.1109/WCAE53984.2021.9707614}
\showDOI{\tempurl}


\bibitem[Ringer(2021)]%
        {blog1}
\bibfield{author}{\bibinfo{person}{Talia Ringer}.}
  \bibinfo{year}{2021}\natexlab{}.
\newblock \bibinfo{title}{Introducing {SIGPLAN-M}}.
\newblock
\newblock
\urldef\tempurl%
\url{https://blog.sigplan.org/2021/01/05/introducing-sigplan-m/}
\showURL{%
\tempurl}


\bibitem[Volunteers(2022a)]%
        {guidelines}
\bibfield{author}{\bibinfo{person}{{SIGPLAN-M} Volunteers}.}
  \bibinfo{year}{2020-2022}\natexlab{a}.
\newblock \bibinfo{title}{Guidelines for Long-Term Mentorship}.
\newblock
\newblock
\urldef\tempurl%
\url{https://docs.google.com/document/d/1eueaiHjNhhqb3JC3wnNlTRcFquNl7BN1OqBczUBumNU/edit}
\showURL{%
\tempurl}


\bibitem[Volunteers(2022b)]%
        {website}
\bibfield{author}{\bibinfo{person}{{SIGPLAN-M} Volunteers}.}
  \bibinfo{year}{2021-2022}\natexlab{b}.
\newblock \bibinfo{title}{{SIGPLAN-M}}.
\newblock
\newblock
\urldef\tempurl%
\url{https://www.sigplan.org/LongTermMentoring/}
\showURL{%
\tempurl}


\bibitem[Wickerson(2022)]%
        {blog2}
\bibfield{author}{\bibinfo{person}{John Wickerson}.}
  \bibinfo{year}{2022}\natexlab{}.
\newblock \bibinfo{title}{People of {PL}: Special Mentoring Edition}.
\newblock
\newblock
\urldef\tempurl%
\url{https://blog.sigplan.org/2022/06/13/people-of-pl-special-mentoring-edition/}
\showURL{%
\tempurl}


\end{thebibliography}
